\documentclass[conference, preprint]{IEEEtran}
\IEEEoverridecommandlockouts
\usepackage{cite}
\usepackage{amsmath,amssymb,amsfonts}
\usepackage{algorithmic}
\usepackage{algorithm}
\usepackage{graphicx}
\usepackage{textcomp}
\usepackage{xcolor}
\usepackage{comment}
\usepackage{todonotes}
\usepackage{authblk}
\def\BibTeX{{\rm B\kern-.05em{\sc i\kern-.025em b}\kern-.08em
    T\kern-.1667em\lower.7ex\hbox{E}\kern-.125emX}}
\begin{document}

\title{Accelerating Bidiagonalization of Banded Matrices through Memory-Aware Bulge-Chasing on GPUs}

%\author{%\IEEEauthorblockN{ Anonymous Author(s)}

\author[1,3]{Evelyne Ringoot}
\author[1,2,4]{Rabab Alomairy}
\author[1,5]{Alan Edelman}

\affil[1]{Computer Science \& Artificial Intelligence Laboratory, and Department of Mathematics, \protect \\Massachusetts Institute of Technology, USA.}
\affil[2]{Computer, Electrical and Mathematical Sciences and Engineering Division, \protect \\
King Abdullah University of Science and Technology, KSA.}

\affil[3]{\textit 
{eringoot@mit.edu}\quad$^{4}$\textit
{rabab.alomairy@mit.edu}\quad$^{5}$\textit {edelman@mit.edu}
}

\maketitle

\begin{abstract}

The reduction of a banded matrix to bidiagonal form is a critical step in the calculation of Singular Values, a cornerstone of scientific computing and AI. Although inherently parallel, this step has traditionally been considered unsuitable for GPUs due to its memory-bound nature. However, recent advances in GPU architectures, such as increased L1 memory per Streaming Multiprocessor (SM)/Compute Unit and larger L2 caches, have shifted this paradigm. In this work, we present the first GPU-accelerated algorithm for reducing a banded matrix to bidiagonal form, integrated into an open-source software package. Our algorithm builds on prior multicore CPU cache-efficient bulge-chasing methods, adapted to modern GPU architectures to optimize throughput. Leveraging Julia’s high-level array abstractions and KernelAbstractions.jl, we implement a single function that is both hardware-agnostic and data-precision-aware, running efficiently across NVIDIA, AMD, Intel, and Apple Metal GPUs. We develop a hardware-aware performance model to guide tuning and identify key hyperparameters that govern optimal GPU performance for memory-bound workloads. We show that such workloads, when carefully optimized, can achieve substantial speed-ups on modern GPUs: our implementation outperforms multithreaded CPU libraries (PLASMA, SLATE) starting from matrix sizes as small as $1024\times1024$, and achieves over 100$\times$ speed-up on $32k\times32k$ matrices. Moreover, the algorithm’s performance scales linearly with the matrix bandwidth, enabling efficient reduction of matrices with larger bandwidths, previously considered impractical. 
\end{abstract}

%© {Ringoot, Alomairy, Edelman } {2025}. 
%This is the author's pre-print version of the work. It is posted here for your personal use only.
\section{Introduction}
The singular value decomposition (SVD) 
\textcolor{black}{decomposes a matrix $A$ into the product $A=U \Sigma V^T$, where $U$,$V$ are orthogonal matrices and $\Sigma$ is a diagonal matrix.} It 
is a fundamental numerical tool in a wide range of applications, underpinning scientific computing, machine learning, and data analytics\cite{Martin01122012}. 
\textcolor{black}{In recent work, singular values have played a key role in for example quantum information processing,\cite{PhysRevResearch.6.043227}, while the full decomposition—including the unitary matrices—has been central to Low-Rank Adaptation (LoRA) in large language models \cite{hu2022lora,NEURIPS2024_ed9f00cb}.}
While dense singular value algorithms have been extensively studied, they still suffer from major scalability challenges. 
\textcolor{black}{With the rise of GPUs, their performance for large data has improved significantly, thanks to the hierarchical massively parallel nature of GPUs. }
On modern, large-scale hardware, the SVD is typically calculated in a three-stage process: reduction from dense to banded form, reduction from banded to bidiagonal form (so-called bulge-chasing), and bidiagonal reduction to diagonal. The three-stage process increases computational density: it has been shown to improve the SVD performance on GPUs by a factor of 10 versus the direct dense-to-bidiagonal algorithm~\cite{gates2018accelerating}.
Research into the implementation of the compute-bound first stage\cite{doi:10.1137/17M1117732} and the third stage\cite{10.1145/1693453.1693472, 10.1145/3716171,6468510,10.1145/2482767.2482813} on GPU devices has been extensive, \textcolor{black}{and resulted in order-of-magnitude speed-ups versus CPU implementations\cite{gates2018accelerating}. }
\textcolor{black}{Previous work has also shown that the multi-GPU communication-avoiding algorithms can be adapted for single-GPU optimization and can be the basis for hardware-agnostic libraries performing on par with optimized vendor libraries, using the Julia Language abstraction~\cite{ringoot2025performant}.}
Meanwhile, the second stage remains underexplored on modern GPUs. In particular, a decade ago, Dongarra~\cite{dongarra2014accelerating} and Gates\cite{gates2018accelerating} remarked that ``accelerators perform poorly when dealing with memory-bound fine-grained computational
tasks (such as bulge chasing)". 
\textcolor{black}{Additionally, banded matrices occur not only in the SVD of large dense matrices, but equally directly in applications such as spectral methods for partial differential equations\cite{meuris2023machine}.}

\textcolor{black}{While GPUs offer hierarchical massive parallelism that far exceeds classical CPUs, this advantage comes with significant challenges—particularly in managing memory across multiple levels of the hierarchy and balancing synchronization overhead against parallel efficiency. Historically, the limited capacity of low-level GPU memory (L1/L2 caches) has led to a preference for CPU-based, cache-optimized algorithms when addressing memory-bound computations. } 
\textcolor{black}{Recent advancements in GPU architecture, for example, NVIDIA's Hopper architecture's increased L1 and L2 cache size and bandwidth ~\cite{h100_pv_mi250}, and AMD's Infinity Memory Layer\cite{amd_mi300_1}, have significantly shifted the memory-compute balance and have changed the long-held belief that memory-bound algorithms cannot be optimized for GPUs. In particular, larger and faster memory at the lowest levels enables fast implementations of memory-bound algorithms,
opening the door for memory-bound algorithms to evolve into high-throughput GPU kernels.
In this work, we revisit the long-overlooked memory-bound banded-to-bidiagonal reduction kernel—a critical component of the singular value pipeline—across varying matrix bandwidths, re-evaluating its suitability for GPU acceleration in light of recent architectural advances. 
\begin{enumerate}
    \item \textbf{We introduce the first fully GPU-resident memory-aware bulge-chasing algorithm for reducing a banded matrix to bidiagonal form.} We detail the memory-aware algorithmic design and synchronization trade-offs that enable outperforming optimized established CPU-based implementations by a factor of 10-100.
    \item \textbf{We introduce a cache-efficient bulge-chasing strategy for large matrix bandwidths through successive bandwidth reduction.} 
    The bandwidth is successively reduced by a set number of bandwidth reduction elements to take optimal advantage of GPU cache levels, even at large bandwidths, significantly increasing the bandwidths at which bulge-chasing is performant. 
    \item \textbf{We provide this functionality in an open-source, hardware-agnostic, and data precision-agnostic library.} 
    Our library provides single, half, and double precision across NVIDIA, AMD, Intel, and Apple GPUs, covering 
    consumer and integrated GPUs, without relying on vendor libraries, enabled through the Julia programming language, which provides meta-programming and type-inference abilities that enable cross-hardware and cross-precision performance through a single implementation~\cite{besard2019,kajl} that is optimized at compile time.     
    We examine the performance across different architectures to conclude that L1 and L2 bandwidth are the main performance determinants, highlighting the importance of algorithmic considerations for future GPU hardware design.
\end{enumerate}
}
 %Our approach demonstrates that, with appropriate algorithmic reformulation, this memory-bound kernel, historically limited to the CPU, can now be efficiently executed on recent GPU architectures without resorting to fallback CPU routines, host-device synchronization, or lossy approximations. 
  \textcolor{black}{The strategy and discussion presented here could, in the future, serve as a template for adapting other memory-bound algorithms to GPUs.
  %—particularly as newer architectures provide larger L1/L2 caches, additional cache hierarchy levels, and increased memory bandwidth.
  }
Accelerating this stage can also benefit the full SVD pipeline. First, our GPU algorithm—while currently focused on singular values—can be extended to include singular vectors. The back-transformation step (i.e., computing singular vectors from the bidiagonal form) remains a major bottleneck in GPU-resident SVD algorithms \cite{10.1145/3721145.3730411,10.1145/3710848.3710894}. 
Second, our algorithm scaling efficiently with increasing matrix bandwidth opens up the possibility of rebalancing the traditional trade-off in singular value pipelines—between faster reduction to banded form (which favors larger bandwidths) and faster reduction to bidiagonal form (which favors smaller bandwidths)
%. By accelerating the second stage, our work could shift this trade-off, enabling larger bandwidths 
and improving end-to-end performance.

\section{Background}
\subsection{Related work}
\subsubsection{Singular value solvers}
While three-stage full SVD algorithms for dense matrices (also referred to as two-stage SVD in the literature) that pass through an intermediate banded and bidiagonal form are well established, the second stage—reduction from banded to bidiagonal form—remains under-optimized in modern accelerator-rich environments. Early work on CPU implementations focused on exploiting the memory hierarchy via blocked Householder transformations, bulge-chasing algorithms, and successive band reduction~\cite{dongarra2019plasma,gates2018accelerating,ballard2012communication,10.1145/2382585.2382587}. 
These designs, implemented in libraries such as PLASMA~\cite{haidar2011parallel}, SBR~\cite{10.1145/365723.365736}, FLAME,~\cite{10.1145/2535371}, ELPA~\cite{marek2014elpa}, and LAPACK, relied heavily on task-based runtimes and BLAS3 kernels to manage data dependencies. On GPUs, 
most frameworks either offload and optimize only the first stage (dense-to-band) on the GPU, executing the second stage on CPUs, as seen in early hybrid implementations~\cite{gates2018accelerating} and in SLATE~\cite{gates2025evolution,dongarra2014accelerating,10.1145/3178442.3178448}, or bypass the banded intermediate altogether using a one-stage reduction approach, as in MAGMA~\cite{abdelfattah2024magma,haidar2014novel,TOMOV2010645}, rocSOLVER\cite{rocsolver}, early research \cite{10.1145/2535371,5161058, 10.1145/1851476.1851512}, and more recent efforts~\cite{liu2025efficient}. 
\textcolor{black}{The former three-stage approach was shown to be up to ten times more performant than the one-stage approach thanks to improved computational density~\cite{doi:10.1137/17M1117732, gates2018accelerating}, but only if a high-performance CPU is available to execute the second stage.}

A separate body of work focuses on calculating approximate or largest singular values and vectors on GPUs and vectors~\cite{STRUSKI2024123462}; we focus in this work on dense full solvers.
\subsubsection{Eigenvalue solvers}
In contrast to the singular value solvers,
where the direct reduction to bidiagonal is preferred in GPU implementations, there is a large body of work that implements and documents the two-stage solver on hybrid CPU-GPU systems\cite{gates2025evolution,YU2021107808,10.1007/978-3-642-55224-3_63,https://doi.org/10.1002/cpe.5915, 10.1007/978-3-030-43229-4_7}. CPU-based parallelization strategies include an extension of the divide-and-conquer algorithm\cite{doi:10.1137/110823699} or the QR-based bulge-chasing method\cite{haidar2011parallel}.
\textcolor{black}{Multicore CPU implementations are widely available for the symmetric banded eigenvalue case: both open-source libraries AMD rocSolver\cite{rocsolver} and MAGMA\cite{abdelfattah2024magma} include optimized eigenvalue solvers for symmetric banded diagonal and tridiagonal systems, but no singular value solvers for non-symmetric banded diagonal systems.  The NVIDIA cuSOLVER\cite{cusolver} library similarly provides a GPU-resident symmetric banded diagonal eigenvalue solver, but no general banded singular value solver. }
Very recently, more work has emerged on the GPU-resident second stage of the eigenvectors for symmetric matrices and by Zhou et al\cite{10.1145/3721145.3730411}, who still consider the bulge-chasing CPU-resident, and independently by Hansheng et al.~\cite{10.1145/3710848.3710894}, who also include the bulge-chasing. While the latter authors' initial implementation used volatile values for inter-block synchronization, an approach not consistent with the CUDA programming guide\cite{progguide}, the authors have since stated this has been addressed in the new version of their codebase~\cite {emailsvd}. 
\subsubsection{Bidiagonal and tridiagonal singular value and eigenvalue solvers}
Substantial progress has also been made on the third stage of the reduction pipeline—solving tridiagonal or bidiagonal matrices.
 On CPUs, several high-performance parallel implementations have been proposed, including 
 and recursive panel reductions designed to exploit multicore architectures and NUMA-aware memory layouts~\cite{ballard2012communication, haidar2011parallel, ltaief2013high,7161495}, implemented in libraries such as PLASMA~\cite{dongarra2019plasma} and ELPA~\cite{marek2014elpa}. 
 Over the past decade, these algorithms have been implemented and optimized on GPUs as well~\cite{vomel2012divide,yu2021gpu}, including non-vendor-specific implementations~\cite{laszlo2016manycore,tolmachev2025high,miao2015fast}.
 Novel faster algorithms for the third stage are still emerging\cite{6468510,app142210716,doi:10.1137/120876605,10.1145/3361746, 10.1145/2535371}.  In addition, recently, the divide-and-conquer algorithm typically used for the bidiagonal matrix was extended to the reduction of banded matrices to singular values, bypassing the bi- or tridiagonal stage\cite{https://doi.org/10.1002/nla.2365,LIAO20161933,Li2022ParallelStructured,7284415}.
\subsubsection{Research question}
In contrast with ample work on the first and third stages for singular value calculation, research on GPU-based reduction from banded to bidiagonal form has been lagging, led by the belief that memory-bound algorithms cannot be accelerated on GPUs. 
\textcolor{black}{Meanwhile, it has been shown that two-stage algorithms are more performant on GPUs than one-stage algorithms~\cite{doi:10.1137/17M1117732, gates2018accelerating}.}
 Compared to symmetric tridiagonalization, reduction from banded form to bidiagonal form indeed presents greater challenges, 
 \textcolor{black}{
 as illustrated by the integration of GPU-resident two-stage algorithms for eigendecomposition in contemporary libraries, while relying on less performant one-stage algorithms for the singular values\cite{abdelfattah2024magma}. 
 On the GPU, efficient cache line utilization is a critical factor for the performance of memory-bound algorithms that is indeed more complex in the non-symmetric case: due to the linear data layout, symmetric eigenvalue problems benefit from data alignment, while asymmetric matrix transformations require irregular data access, arising from the application of both left and right orthogonal transformations to a matrix, which is aligned and can be accessed in a performant manner, and its transpose, which is not aligned in memory and requries more complex optimizations for performant data access.  }

 In this work, we seek to address this gap in knowledge and revisit the bidiagonal reduction problem in light of recent GPU architectural advancements—particularly increased L1 and L2 cache capacity and bandwidth and propose the first fully GPU-resident implementation of band-to-bidiagonal reduction through bulge-chasing. 
\subsubsection{Bulge-chasing algorithms}
The bulge-chasing algorithm was initially proposed as a parallel algorithm in 1996\cite{LANG19961, lapack} that accelerates the reduction of banded matrices to bidiagonal form. It introduces orthogonal transformations that annihilate non-zero elements. Each annihilated element introduces a new non-zero 'bulge' that must be chased toward the matrix boundary through successive transformations. Research since then has been extremely sparse: in 2012, the scheduling and cache-efficiency was investigated\cite{6267821}, and simultaneously Ballard et al\cite{ballard2012communication} proposed a communication-avoiding strategy grouping more elements together. In contrast, bulge-chasing is also a commonly used algorithm for different use cases: reduction of symmetric banded matrix to tridiagonal for eigenvalue calculation\cite{haidar2011parallel,MANIN2023102998}, and reduction of matrices to upper Hessenberg form\cite{KARLSSON2014271,doi:10.1137/23M1547093,10.1145/2699471}, both of which have a higher degree of compute-intensity, which have been more extensively studied.

\subsection{Julia language GPU abstractions}
 Our approach leverages Julia’s GPU programming ecosystem to deliver a portable, precision-agnostic algorithm capable of targeting multiple GPU architectures efficiently, through type-inference that allows the compiler to optimize generic functions just-in-time (JIT) for optimal performance~\cite{besard2019,ringoot2025performant}. These abstractions allow us to focus on the algorithmic structure while preserving performance portability and fine-grained concurrency. We make use of
 GPUArrays.jl\cite{gpuajl} to abstract hardware-specific memory layout and device allocation, and KernelAbstractions.jl\cite{kajl} to write a single architecture-agnostic GPU kernel that compiles to NVIDIA~\cite{besard2019}, AMD~\cite{AMDGPU}, Intel~\cite{oneapi}, and Apple GPUs~\cite{metal}. 
 Contrary to approaches such as ArrayFire\cite{malcolm2012arrayfire}, which provide a high-level unified API, but on the lower level still rely on vendor libraries such as CUSOLVER and rocSOLVER, the Julia kernel code is translated directly to machine code (e.g., PTX) through the LLVM compiler\cite{carlson2025ccodegenconsideredunnecessary, besard2019}. 
 Such an approach allows straightforward implementation of novel hardware and data types, making the implementation extensible in the future, for example, for novel hardware such as TPUs\cite{doi:10.1073/pnas.2122762119}. 
Importantly, it has been demonstrated that such agnostic implementations do not require sacrificing performance relative to vendor-optimized functions
~\cite{carrica2025toward,dla,ringoot2025performant}. 
Novel open-source libraries such as NextLA.jl~\cite{the-code} have been released, taking advantage of these features to provide user-friendly, generic, scalable linear algebra functionality.

\section{Algorithm}

Singular values are computed by applying orthogonal transformations to subsets of rows and columns, with each transformation designed to annihilate specific off-diagonal elements (i.e., transform them to zero). In the second phase of the banded SVD reduction, these elements are referred to as bulges, and the process of eliminating them is known as bulge chasing.
 In this work, we follow the parallel bulge-chasing strategy originally proposed by Lang~\cite{LANG19961}, and later extended by Haidar et al.~\cite{6267821} and Ballard et al.~\cite{ballard2012communication}, where once a bulge has been chased far enough, the next bulge can be processed in parallel. 
\textcolor{black}{These strategies were previously published for and confined to CPUs, where the highly efficient large caches resulted in high performance. However, the massively parallel nature of GPUs is unmatched by CPUs, resulting in significant performance improvements when the memory levels are adequately exploited.}

\begin{figure}[h]
\centerline{\includegraphics[width=\linewidth]{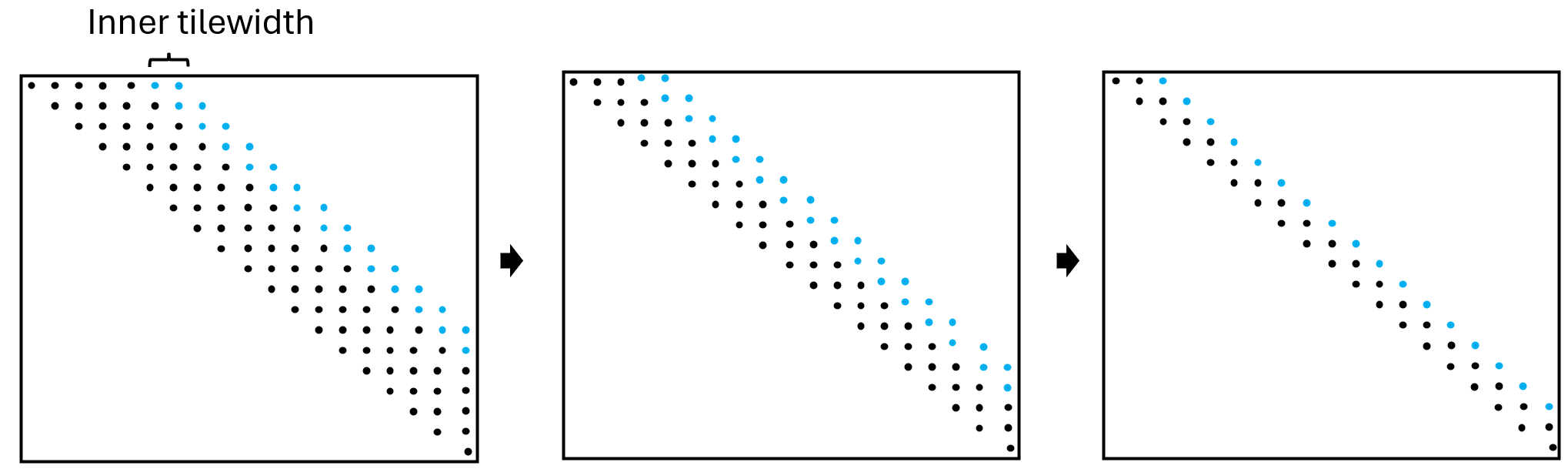}}
\caption{Illustration of the process of successive reduction of the bandwidth of the matrix by inner tilewidth for inner tilewidth 2 and total bandwidth 6. 
%The proposed strategy increases memory locality and cache reuse, crucial for the memory-bound algorithm on the GPU, where limited memory and memory bandwidth are available.
}
\label{fig1}
\end{figure}

 \begin{figure*}[htbp]
\centerline{\includegraphics[width=\linewidth]{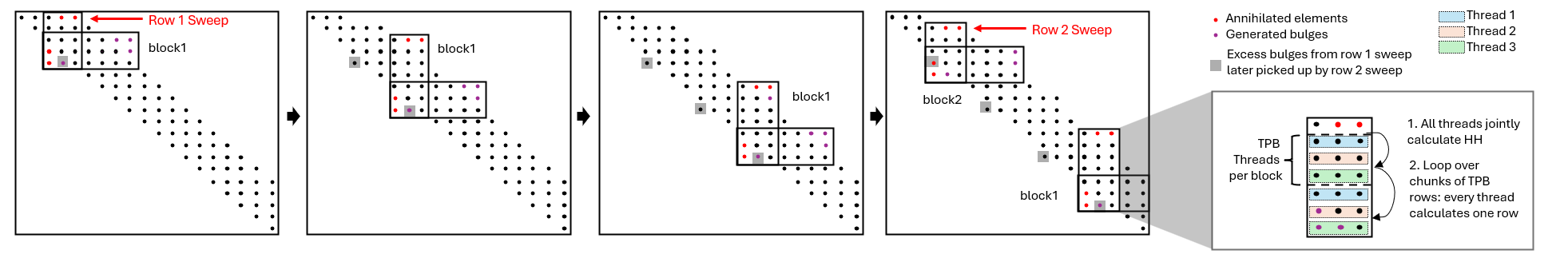}}
\caption{Visualization of the bandwidth-tiled, sweep-based GPU implementation of band-to-bidiagonal reduction using Householder reflectors. Each row sweep (e.g., Row 1, Row 2) processes a bandwidth tile, chasing bulges (purple dots) down the matrix by applying orthogonal transformations to annihilate matrix elements (red dots). Excess bulges generated during one row’s sweep are deferred (gray squares) and picked up in subsequent sweeps. The right panel illustrates how each thread block cooperatively computes transformations across rows using chunked tiling, where each thread handles a row and loops over multiple elements. 
%This design enables fine-grained parallelism, improved memory locality, and GPU-resident execution across bandwidth tiles.
}\label{figalgo}
\end{figure*}
We propose a novel GPU-aware extension to this parallelization strategy. Specifically, we split the matrix into bandwidth tiles (TW) and perform the reduction in successive bandwidth chunks, rather than reducing the entire bandwidth at once, as illustrated in Figure \ref{fig1}. Our work is the first to demonstrate that such CPU-style bulge-level parallelism can be successfully deployed in a fully GPU-resident band-to-bidiagonal reduction, and to our knowledge, the first to treat bandwidth tiling as a main optimization strategy. 
This tiling strategy enables better cache reuse and improved memory locality for the reduction from banded form to bidiagonal form—especially of matrices with large bandwidths: as the GPU cache memory is limited, processing successive bandwidth reduction successively rather than at once retains the processed data at the lowest memory level with lowest latency(L1 and L2).

\textcolor{black}{
Algorithm~\ref{alg:pseudo} presents the pseudocode for our GPU-resident implementation of the band-to-bidiagonal reduction using Householder transformations. The \textbf{outermost loop} (line~1) reduces the matrix bandwidth in stages, each time decreasing it by a fixed inner tilewidth (\texttt{TW}). 
The \textbf{second loop} (line~3) performs a \emph{sweep} across all matrix rows, chasing down a single bulge until it exits the matrix. 
\textcolor{black}{Note that the initial start rows $k_0$ are below zero, but get corrected on line 7: the annihilation of the original bulges is applied to a smaller number of vectors than the following bulges, as illustrated as well in Figure \ref{figalgo}.}
%Each row has a number of elements that is annihilated in a so-called sweep (line 3), introducing  a new bulge that must be chased toward the matrix boundary through successive transformations: the sweep remains attached to the bulges it chases. 
For each row, the algorithm generates a sequence of \emph{row-bulges}, which are processed in the \textbf{third loop} (line~5). Each of these bulges introduces new elements that must be annihilated and may themselves generate additional bulges during the sweep.
The annihilation is executed through finding a proper householder vector and applying it to the respective rows or columns (lines 8-9).  Consecutive sweeps can be executed in parallel, as long as the previous sweeps have progressed far enough (condition on line 6). 
We thus have two levels of hierarchical parallelism in the algorithm: between the different sweeps, and between the application of the parallel vector-vector products of the householder transformations: ideal for parallelization across threadblocks and threads. This cascading process is illustrated in Figure~\ref{figalgo}:  each thread block processes a tile along the diagonal, sweeping row-by-row and applying annihilations of bulges in parallel across threads. 
Excess bulges generated in one sweep are deferred and picked up in subsequent sweeps, enabling fine-grained pipelining across row tiles, as proposed by \cite{ballard2012communication}. As singular values are invariant under orthogonal transformations,  the order of the transformations is indeed interchangeable. }
This process is also illustrated in Figure~\ref{figalgo}, which shows the sweep execution for an inner tilewidth of~2. From left to right, it depicts:
\begin{itemize}
    \item The first row sweep (row~1) with bulge at $k = 1$,
    \item Subsequent bulges at $k = 3$ and $k = 7$,
    \item The final bulge of row~1 at $k = 10$, and the start of the second row sweep (row~2) at $k = 2$.
\end{itemize}
\vspace{0.2in}
 \begin{algorithm}
 \caption{Reduction of a banded matrix to bidiagonal form using Householder vectors (HH)}
 \label{alg:pseudo}
 \begin{algorithmic}[1]
   \renewcommand{\algorithmicrequire}{\textbf{Variable:}}
 \REQUIRE Input Bandwidth = BW$_0$
 \REQUIRE Inner Tilewidth = TW 
 \REQUIRE Matrix size = n
  \FOR {Bandwidth reduction step $i = 1 : $ BW$_0/$TW}
  \STATE{BW$_i=$BW$_{i-1}-$TW }
  \FOR {\textbf{parallel}: sweep $R = 1 : n$}
  \STATE{Start index $k_0 = $R$-$TW$-3($R$-1) \cdot$ BW$_i$}
  \FOR {kernel on row $k=k_0 : BW_i : n$}
  \IF{ $ k \geq R-TW$}
  \STATE If $k=R-TW$, use $k=R$ instead 
  \STATE Calculate HH for row $k$ for annihilating TW elements and apply to rows below 
  \STATE Calculate HH for left-most generated column-bulge annihilating TW elements and apply to columns to the right

  \ENDIF 
  \STATE
  \textbf{synchronize parallel for}
  \ENDFOR
    \ENDFOR
    
  \ENDFOR
 \end{algorithmic} 
 \end{algorithm}

In each subfigure, red elements represent values annihilated by orthogonal transformations, while purple elements represent bulges generated during the sweep. Gray squares indicate bulges left unprocessed by the row~1 sweep, which are deferred and later annihilated during the row~2 sweep.

\subsection{Managing read-and-write dependencies}
To ensure correctness while enabling parallelism, the algorithm enforces a three-cycle separation between sweeps on consecutive rows. 
\textcolor{black}{That is, after every third row-bulge in a sweep completes, the next row can begin its sweep without overlapping data access. %A sweep here refers to the “chasing down” of a row of bulges to the bottom of the matrix. 
Each cycle—or GPU kernel launch—corresponds to the annihilation of two groups of elements: one horizontal and one vertical. Let TW denote the tilewidth (i.e., the bandwidth portion reduced in a single iteration), and BW the total bandwidth before reduction. The reason a three-cycle separation is required becomes apparent when we consider the following:
\begin{itemize}
    \item \textbf{\textit{Every sweep accesses $1 + \text{BW} + \text{TW}$ consecutive elements.}} For the annihilation of the bulges in row $R$, the data accessed in a single cycle spans from row $R$ to $R + \text{BW} + \text{TW}$ vertically, and covers the same number of columns horizontally (Figure~\ref{figalgo}(b)). The initial cycle (Figure~\ref{figalgo}(a)) accesses fewer elements and is thus a special case within the general pattern. The next sweep, for row $R + 1$, begins one row lower and accesses data up to row $R + 1 + \text{BW} + \text{TW}$.
    \item \textbf{\textit{After three execution cycles, the next sweep can begin without data overlap.}} A separation of $n$ execution cycles between consecutive sweeps is sufficient if the last element accessed by the next sweep does not overlap with the first element accessed in the $n$-th cycle of the current sweep, observing that every cycle moves the data access by BW elements forward. This condition can be expressed as $R + n \cdot \text{BW} > R + 1 + \text{BW} + \text{TW}$. This inequality is never satisfied for $n = 1$, and is satisfied for $n = 2$ only when reducing to bandwidths larger than bidiagonal form, i.e., $\text{BW} > 1 + \text{TW}$. In our setting, however, the reduction targets bidiagonal form (i.e., $\text{BW} \ge 1 + \text{TW}$), and the original matrix bandwidth is always larger than bidiagonal (i.e., $\text{BW} > 1$). Therefore, a separation of $n = 3$ execution cycles is always sufficient to guarantee non-overlapping data access between consecutive sweeps.
\end{itemize}
This enables concurrent execution across rows without violating data dependencies. Each bulge-chasing step (lines~6--9) is implemented as a GPU kernel, with each parallel sweep (line~3) mapped to a separate thread block. Between iterations of the bulge-chasing steps, the GPU device performs a synchronization step (line~10) by completing the current kernel and launching the next, ensuring the correct propagation of transformations across the matrix. 
}

  \begin{algorithm}
 \caption{Memory-Aware GPU Kernel for Row-Bulge Execution with TPB Threads per Block.}
 \label{alg:gpu}
 \begin{algorithmic}[1]
   \renewcommand{\algorithmicrequire}{\textbf{Input}}
   \renewcommand{\algorithmicensure}{\textbf{Definition}}
   \ENSURE HH(X): calculate Householder vector of X in place
   \ENSURE HH(X,Y): Apply householder vector X to Y
 \REQUIRE Banded matrix A
 \REQUIRE Row-bulge $k$ to chase
 \REQUIRE Inner tilewidth TW
 \REQUIRE Current bandwidth BW$_0$= BW$_1$+TW
 \STATE \textbf{Thread memory:} $A_i$ (TW+1)
 \STATE \textbf{Block Memory:} $X$ (TW+1)
  \STATE $\forall$ threads in block cooperatively: $X \leftarrow $ A[k,..]
  \STATE synchronize threads
  \STATE $\forall$ threads in block cooperatively: HH($X$)
   \STATE $\forall$ threads in block cooperatively: A[k,..]$ \leftarrow X $ 
  \STATE synchronize threads
  \FOR{$l: 0 \rightarrow (BW_0+TW)/TPB-1$}
  \STATE $\forall$ Thread $i:$ Row to calculate $r=k+l\cdot CPB+i$
\STATE{ \qquad \qquad  \quad $A_i \gets A[r,...] $ }
\STATE{ \qquad \qquad  \quad HH($X$,$A_i$) }
\STATE{ \qquad \qquad  \quad $ A[r,...] \gets A_i $ }
  \ENDFOR
    \STATE synchronize threads
    \STATE Apply procedure above 3-12 for the associated column-annihilation
 \end{algorithmic} 
 \end{algorithm}

\subsection{Memory-Aware GPU Kernel Implementation}
We combine bulge-level concurrency, bandwidth tiling, and multi-sweep bulge management into a unified GPU kernel. 
 Since bulge chasing is primarily memory-bound, maximizing performance requires careful tuning of memory hierarchy usage rather than raw compute throughput. 
For the GPU implementation of this memory-bound kernel, we prioritize low-level memory bandwidth optimization over peak compute throughput. Specifically, we limit occupancy—i.e., the number of concurrent threads—to reduce register pressure. 
The memory access strategy is outlined in Algorithm~\ref{alg:gpu}. For clarity, index calculations and zero-condition handling are omitted. Details of the Householder reflector computation and the treatment of near-zero elements are implemented according to prior work on tile-QR decomposition~\cite{ringoot2025performant}. In Algorithm~\ref{alg:gpu}, the elements to be annihilated, along with one element to their left, are first cooperatively loaded by all threads into shared memory (line 3), stored in L1. Once synchronized (line 4), the threads jointly compute the corresponding Householder reflector (line 5), which is then written back into the original memory location (line 6). A second synchronization ensures a consistent state before applying the transformation. Next, the threads loop over chunks of the target row block (lines 8–13). Each thread loads a row of width \texttt{TW+1} into register memory (line 10), applies the previously computed Householder transformation (line 11), and writes the updated data back to global memory (line 12). 
\textcolor{black}{Register values are kept in the lowest memory level up until capacity, at which point the compiler spills them into L2 memory, a lower-latency cache(in CUDA 12.8).}
This chunking strategy is shown on the right of Figure \ref{figalgo}, where threads cooperate to apply the transformation to subsets of the matrix. Finally, the same procedure is repeated within the same GPU kernel to annihilate the leftmost column of the newly generated bulge (line 15). In the early iterations of each row sweep (small $k$), slight adaptations are applied, since fewer rows exist below the annihilated row—this edge case is illustrated in the left panel of Figure\ref{figalgo}. 

While this strategy may result in register spills that are cached in L2 for the individual rows and columns held by threads, it remains more beneficial than relying on L1 shared memory for storing individual vectors to process. First, due to the limited capacity of shared memory, using L1 would significantly restrict parallelism, as the number of concurrently processed rows or columns would be constrained—though this can be tuned based on the hardware architecture. Second, shared memory bandwidth is considerably lower than register access bandwidth, and retaining a subset of values in registers contributes directly to the high performance of the kernel. Third, this approach reduces the need for inter-thread synchronization, further enhancing performance.
By processing only a subset of bulges at a time—\texttt{TPB} rows of width \texttt{TW+1}—we keep register usage bounded and tunable depending on L2 memory bandwidth. 
Finally, future architectural developments such as shared memory space for register spilling and shared memory could prove to speed up the current algorithm even further. We currently use CUDA 12.8 for benchmarking, whereas this feature is now included in CUDA 13.0\cite{registerspills}, \textcolor{black}{which is automatically integrated through the Julia language abstractions when the hardware interface \texttt{CUDA.jl} is updated.}

\subsection{Tunable Parameters and Performance Trade-offs}
The algorithm exposes three key tunable parameters that can be adjusted based on hardware architecture and data precision, enabling high performance across a wide range of GPU platforms. We outline below the performance considerations associated with each. Every threadblock stores one vector of length TW in shared memory (L1), and TPB number of vectors of length TW in thread registers, which the compiler automatically spills into L2 memory when exceeding register memory.
\paragraph{Threads per Block (TPB)} This parameter controls the trade-off between increased parallelism and register/cache pressure within a thread block. Each GPU thread processes a single row or column: if the number of TPB is lower than the number of rows/columns to be processed, the thread processes multiple rows or columns sequentially. Increasing the number of threads reduces this sequentiality and improves parallelism, but increases register usage and pressures the L2 cache (for local vectors spilled from registers). TPB determines the trade-off between parallel throughput and memory utilization.
\paragraph{Inner Tilewidth (TW)} The inner tilewidth determines how much the matrix bandwidth is reduced in a single step. Smaller TW values reduce the length of the householder vector, and vectors to apply it to, and thus memory pressure on L1 and L2 caches by limiting the number of active values per kernel. However, lower TW values also reduce cache line utilization and increase the number of sequential kernel invocations required to reach bidiagonal form. TW must therefore be tuned to trade off kernel granularity and memory reuse efficiency.
\paragraph{Maximum Blocks} Limiting the number of concurrently active blocks per execution unit (e.g., SM, CU, or Xe Core) can improve data locality and reduce memory contention. The \texttt{Max blocks} parameter enforces this limit, enabling each block to access more L1 memory exclusively and reducing L2 cache pressure. However, fewer active blocks may reduce parallelism and thus total throughput. When the number of required bulge-chasing blocks exceeds this limit, we apply software-level loop unrolling, assigning multiple tasks to a single block to be executed sequentially in the same kernel launch. This reduces occupancy in favor of better memory reuse, especially on architectures with limited per-execution unit L1.
For all three parameters, careful tuning based on architectural characteristics is essential to achieve optimal performance.
\vspace{0.2in}

\subsection{GPU Occupancy Modeling for Bandwidth-Tiled Kernels}
\label{sect:occ}
Algorithm~\ref{alg:pseudo} reveals that in the initial stages of the reduction, the number of launched GPU blocks is lower than the number of available execution units. Consequently, our algorithm achieves \textcolor{black}{full hardware utilization} when the matrix has a sufficiently large size-to-bandwidth ratio. In particular, the spacing between bulge-chasing blocks is $3 \cdot \texttt{CBW}$, where \texttt{CBW} is the current bandwidth. The algorithm saturates all GPU resources at matrix size $n$:
\begin{equation}
    \frac{n}{3 \cdot \texttt{CBW}} \geq \texttt{ALUs}
\end{equation}

Table~\ref{tab:occupancy} lists the matrix sizes required to achieve full occupancy on representative GPU architectures for $\texttt{CBW} = 32$, illustrating how hardware characteristics directly influence performance and scalability. 

\begin{table}[ht]
\centering
\caption{Matrix size $n$ required for full GPU occupancy with current bandwidth $\texttt{CBW} = 32$.}
\footnotesize
\label{tab:occupancy}
\begin{tabular}{|l|c|c|}
\hline
\textbf{GPU Architecture} & \textbf{Execution Units (ALUs)} & \textbf{ $n \geq 3 \cdot \texttt{CBW} \cdot \texttt{ALUs}$} \\
\hline
NVIDIA H100      & 132 SMs $\times$  & $50{,}688$ \\
      & 4 warp schedulers = 528 & \\
AMD MI300X       & 304 Compute Units                        & $29{,}184$ \\
Intel PVC Max 1100 & 56 Xe Cores                             & $5{,}376$  \\
\hline
\end{tabular}
\end{table}
\begin{table*}[t]
\caption{Hardware used for benchmarking}
\label{tab:hardware}
\footnotesize
\centering
\begin{tabular}{rccccccc}
\hline
\textbf{}           & \multicolumn{3}{c}{\textbf{NVIDIA}}              & \multicolumn{2}{c}{\textbf{AMD}}  & \textbf{Intel}    & \textbf{Apple} \\ \hline
\textbf{GPU}        & \textbf{A100} & \textbf{H100} & \textbf{RTX4060} & \textbf{MI250X} & \textbf{MI300X} & \textbf{PVC 1100} & \textbf{M1}    \\ \hline
L1/ SM (KB)         & 192           & 256           & 128              & 16              & 32              & 512.0             & 128              \\ \hline
L2  (MB)     & 40            & 50            & 32               & 4               & 256 (L2.5)            & 108               & 12              \\ \hline
Bandwidth (TB/s)    & 2             & 3.35          & 0.28             & 3.2             & 5.3             & 1.2               &  67 GB/s            \\ \hline
L1 latency (cycles) & 40            & 30            & N.A.             & 120             & 120            & 60                &   N.A.             \\ \hline
L2 latency (cycles) & 200           & 300           & N.A              & 230             & 200            & 420               &   N.A.             \\ \hline
ALUs                & 108$\times$4         & 128$\times$4         & 24 $\times$4              & 220             & 304             & 56                & 8$\times$16$\times$8               \\ \hline
Memory (GB)         & 80            & 80            & 8                & 128             & 192             & 48                &   8-16        \\ \hline
Clock Boost (GHz)   & 1.41          & 1.785         & 2.46             & 1.7             & 2.1             & 1.55              &    1.27            \\ \hline
\end{tabular}

\end{table*}

However, in practice, the GPU implementation outperforms state-of-the-art large-scale CPU libraries already at relatively small matrix sizes,
\textcolor{black}{
i.e., much below optimal occupancy (starting at matrix sizes $1024 \times 1024$). }
This is enabled by the dense hierarchical parallelism offered by modern GPUs. At the device level, multiple thread blocks can be launched concurrently across streaming multiprocessors, and within each block, many threads each perform fine-grained vector operations. While CPUs also support hierarchical parallelism via multithreading and SIMD instructions, the scale and efficiency of GPU threading
%—combined with high memory bandwidth and fast shared memory—
enable superior performance for highly parallel memory-bound kernels such as bulge chasing. \textcolor{black}{
 As a consequence, the matrix size can be further increased without significantly increasing the computation cost of the parallel sections, up to full occupancy. Of course, larger matrix sizes still increase the number of sequential sections.}

\subsection{Kernel Profiling}\label{sect:prof}

\begin{table}[h]
\centering
\caption{Kernel profiling on RTX4060 for varying hyperparameters. 
Results show runtime and memory throughput across DRAM, L1, and L2, highlighting the role of cache utilization in performance.
}

\label{tab}
\scriptsize
\begin{tabular}{|r |cccccccc|}
\hline
\textbf{Configuration} &  & &  & \textit{best}  & \textbf{A}  &   \textbf{B}  & & \\ 
\textbf{Threads per block} & \textbf{64} & \textbf{64} & \textbf{32} & \textbf{32}  & \textbf{16}  & \textbf{32} & \textbf{32}  & \textbf{64} \\ 
\textbf{Max blocks}        & \textbf{48} & \textbf{96} & \textbf{96} & \textbf{192} & \textbf{192} & \textbf{96} & \textbf{192} & \textbf{96} \\ 
\textbf{Inner tilewidth}   & \textbf{32} & \textbf{32} & \textbf{32} & \textbf{32}  & \textbf{32}  & \textbf{16} & \textbf{16}  & \textbf{16} \\ \hline
time (us)                  & 147         & 118         & 120         & \textbf{107}          & 124          & 69          & 56           & 70          \\ \hline
{throughput (\%)}     &          &         &       &         &           &          &        &      \\ 
memory   & 38          & 48          & 46          & \textbf{52}           & 45           & 33          & 42           & 34          \\ 
DRAM     & 11          & 14          & 13          & \textbf{16}           & 13           & 13          & 15           & 12          \\ 

L1        & 47          & 59          & 57          & \textbf{64}           & 57           & 41          & 52           & 41          \\ 
L2      & 38          & 48          & 46          & \textbf{51}           & 45           & 33          & 42           & 34          \\ \hline
compute    &          &         &       &         &           &          &        &      \\ 
throughput (\%)    & 12          & 15          & 13          & \textbf{13}           & 23           & 13          & 14           & 14          \\ \hline
warps per sm               & 3.98        & 7.6         & 3.8         & \textbf{6.5}          & 6.69         & 3.84        & 6.63         & 7.57        \\ 
\hline
\end{tabular}
\end{table}

Table~\ref{tab} presents the key kernel profiling metrics as a function of the three hyperparameters: \texttt{Max blocks}, \texttt{Threads per block}, and \texttt{Inner tilewidth}. The measurements were collected on an RTX4060 GPU \textcolor{black}{using NSight Compute directly on Julia code,} using a matrix of size $32k \times 32k$ with $k = 1024$, targeting a reduction of the bandwidth from 64 to 32 or from 64 to 48 so that full parallelism is achieved. We observe that our algorithm makes extensive use of the lower levels of the memory hierarchy (L1 and L2 caches), and that the runtime correlates more strongly with total memory throughput than with DRAM throughput alone. 
Two representative configurations highlight this effect. \textbf{Configuration~A} corresponds to (Threads=16, Max Blocks=192, Tilewidth=32), and \textbf{Configuration~B} corresponds to (Threads=32, Max Blocks=96, Tilewidth=16). Both configurations exhibit similar DRAM throughput, yet Configuration~B shows noticeably lower L1/L2 throughput. Because the smaller tilewidth in Configuration~B annihilates only half as many elements per execution, it would need to run twice to achieve the same reduction as Configuration~A. Consequently, Configuration~A is significantly more performant overall. These observations confirm that effective utilization of all memory levels—especially the L1/L2 memory—is essential for optimal kernel performance. In general, cache size, reuse distance, and latency are key determinants of efficiency for memory‑bound algorithms such as bulge chasing. \textcolor{black}{We highlight the most performant configuration in bold: its runtime over its \texttt{Inner Tilewidth} is the lowest. Configurations with half the \texttt{Inner Tilewidth} need to run twice as often in the overall pipeline, making the configuration with runtime $107 ms$ the fastest overall.}

For reference, we also profiled the \texttt{CUBLAS} matrix‑add kernel \texttt{geam} $B=A+A^T$, executed on $16k \times 16k$ matrices. The \texttt{geam} kernel reaches approximately $78\%$ of DRAM and total memory throughput but only about $18\%$ of L1 and L2 throughput. We compare against this transpose‑add operation because half of our kernel invocations operate along rows and half along columns, giving a fair correspondence in access patterns. While \texttt{geam} achieves higher DRAM utilization, it does not reuse data across warps or blocks, whereas our kernel reuses the same elements multiple times through L1/L2 caching. The slightly lower DRAM throughput we report thus reflects the additional time spent exploiting the intra‑kernel data reuse rather than inefficiency. Although overlapping data transfers between memory levels could further increase throughput, this would also enlarge the required per-block low-level memory size. Given that on‑chip memory is a limiting factor, such additional concurrency would yield poorer cache utilization and ultimately lower performance. Finally, the number of warps per SM is automatically constrained by our high register usage: occupancy is effectively self‑regulated by resource requirements, making the \texttt{Max blocks} parameter implicitly enforced by the hardware.

\begin{figure*}[t]
\centerline{\includegraphics[width=0.955\linewidth]{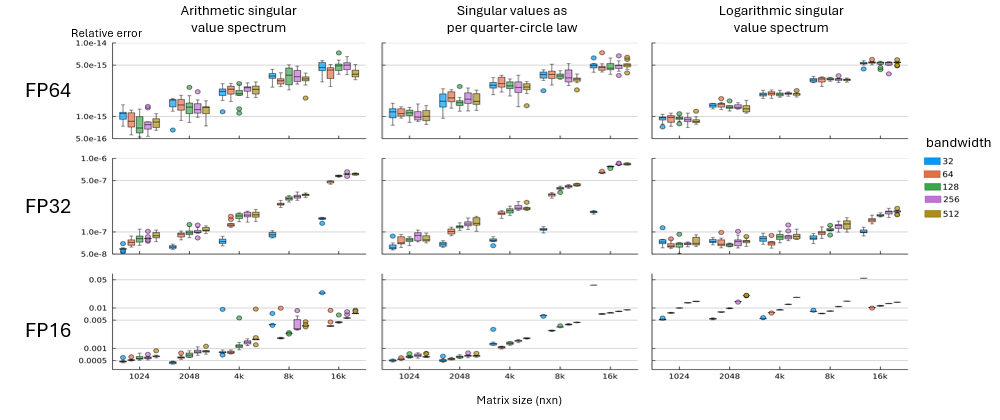}}
\caption{Relative error of singular values computed via GPU-based reduction to bidiagonal form, followed by LAPACK BDSDC in double precision. The boxplots show errors over 10 trials for each singular value profile and precision (FP64/FP32/FP16) across matrix sizes and bandwidths. Results confirm stable accuracy with increasing bandwidth and only moderate error growth with size.}
\label{fig:acc}
\end{figure*}

\section{Evaluation methods}
\paragraph{Selection of hyperparameters} Hyperparameters for the GPU implementation are tuned through a brute-force search (3 parameters across 3-5 values) by hardware architecture(Figures \ref{fig:acc}, \ref{fig:arch}, \ref{fig:slate}, and \ref{fig:hardware}), except where hyperparameter variation is discussed in Figure \ref{fig:parvarfig}.
\paragraph{Storage format}For matrices beyond GPU memory, we use a classical column-major banded storage format, so that we store only the band and the bulges in a matrix of height of the matrix bandwidth, increased by twice the inner tilewidth, and width equal to the original matrix size. 
\paragraph{CPU hardware} For PLASMA(v25.5.27) and SLATE(v2024.10.29) benchmarking, an Intel Xeon Platinum 8462Y+ 32-core CPU with 64 threads was used. \textcolor{black}{A single CPU node is used.}
\paragraph{GPU hardware} The NVIDIA\cite{h100_10,h100_pv_mi250,nvidia_a100_1,nvidia_h100_1,nvidia_a100_2,nvidia_h100_2,a100_10,h100_10, a100_11,chipsandcheese2023nvidiah100}, AMD\cite{amd_mi250_1,amd_mi300_1,h100_pv_mi250,tee2025mall,chipsandcheese2023mi300x}, Intel\cite{intel_pvc_1,intel_pvc_2,intel_pvc_3,h100_pv_mi250} and Apple~\cite{hubner2025apple} hardware characteristics used for benchmarking are shown in Table \ref{tab:hardware}. 
\textcolor{black}{\paragraph{Library compilation} PLASMA and SLATE were built through CMake, in correspondence with the example build files in their GitHub repositories, relying on MKL libraries and 64-bit addressing. Default library parameters were used for testing, except for excluding singular vector calculation.}

\section{Results}
\subsection{Numerical Accuracy}
We assess the numerical accuracy of our algorithm by constructing synthetic matrices with known singular values. Specifically, we generate two random unitary matrices $U$ and $V$ and a diagonal vector $\Sigma$ of prescribed singular values, then form the matrix $A = U \Sigma V^T$. This matrix is first reduced to banded form using the classical block Householder reduction for SVD 
\textcolor{black}{in double precision}
—an approach that has previously shown to preserve high numerical fidelity \cite{ringoot2025performant}. From the banded form, we apply our GPU bidiagonal reduction in reduced precision, and then use the double-precision LAPACK routine BDSDC (bidiagonal divide-and-conquer) to compute the singular values. 
\textcolor{black}{Executing only the banded to bidiagonal in reduced precision ensures no losses of precision from other phases are aggregated.}
We report the relative error between these computed singular values and the original ground-truth values from $\Sigma$, \textcolor{black}{i.e., the ratio of the norm of their difference over the norm of the ground-truth}. Figure~\ref{fig:acc} shows the distribution of relative errors across 30 random instances per data precision: FP64, FP32, and FP16. Each test set includes 10 matrices with an arithmetic singular value spectrum (uniform spacing), 10 with a logarithmic decay, and 10 following a quarter-circle distribution—emulating typical spectrums observed in structured, ill-conditioned, and random matrices, respectively. Singular values are sampled in the $[0, 1]$ interval, but the analysis generalizes to scaled domains via simple normalization. As shown, the algorithm maintains stable accuracy across a range of matrix bandwidths and remains well within acceptable numerical limits for each precision. FP64 errors remain near machine epsilon; FP32 shows a predictable, size-dependent increase; and FP16 retains acceptable relative accuracy for matrices up to $16k \times 16k$, despite the reduced mantissa precision, confirming numerical stability empirically. As expected, we find that the main predictor for accuracy in reduced precision is the singular value profile: lower precision is best suited for well-behaved matrices.
\textcolor{black}{Moreover, we find that increasing the bandwidth while keeping the inner tilewidth constant does not have a substantial impact on performance, confirming that the proposed successive band reduction indeed does not negatively impact accuracy. }

\begin{figure*}[h]
\centerline{\includegraphics[width=0.98\linewidth]{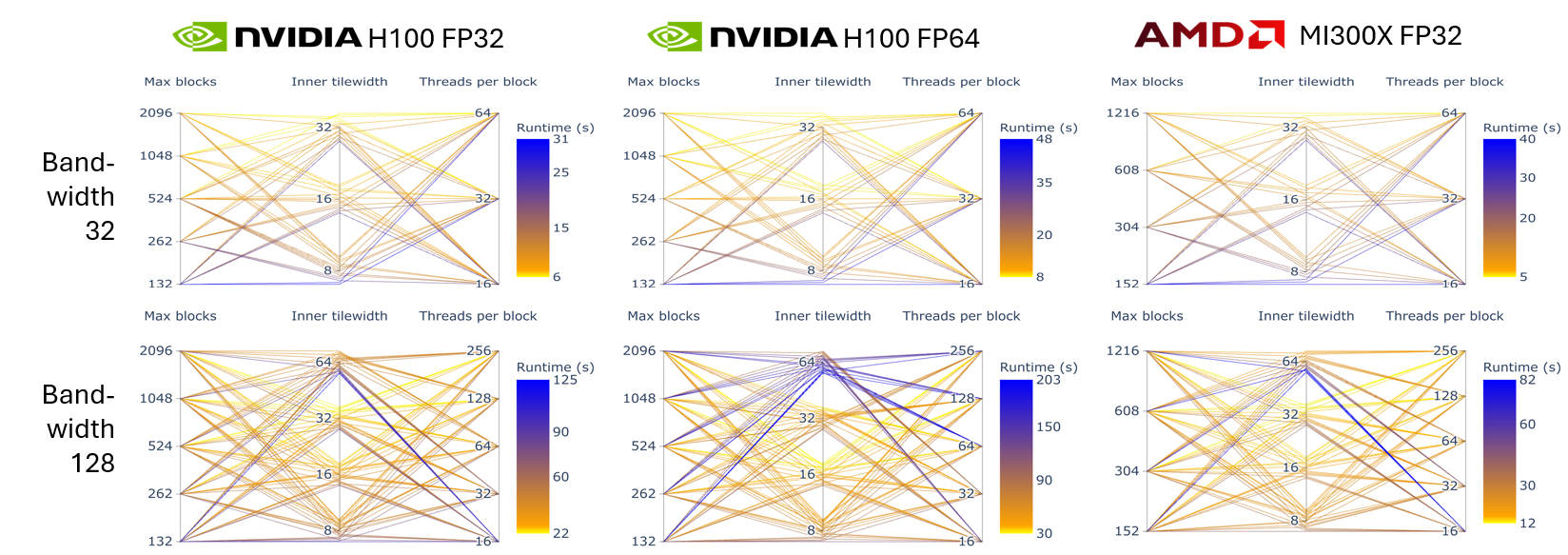}}
\caption{ Performance of the GPU-based reduction algorithm from banded to bidiagonal form across kernel hyperparameters, hardware, and precisions. Parallel coordinate plots show color-coded relative runtimes for combinations of three hyperparameter max blocks, tile width, and threads per block on NVIDIA H100 (FP32, FP64) and AMD MI300X (FP32), at bandwidths 32 and 128. Yellow lines indicate faster configurations. While tile width—optimally matching cache line sizes (32 for FP32, 16 for FP64)—has the greatest impact, both max blocks and threads per block also significantly influence performance, especially at wider bandwidths.}
\label{fig:parvarfig}
\end{figure*}
\subsection{Hyperparameter Tuning Across Devices and Precisions}
\label{sect:paramtune}
The performance of the GPU-accelerated reduction to bidiagonal form for a $65k \times 65k$ matrix is visualized in Figure~\ref{fig:parvarfig} using a parallel coordinates plot~\cite{glendenning2016parameter}; for AMD at bandwidth 128, a $32k \times 32k$ matrix is shown. This plot style is commonly used to evaluate the effect of multiple hyperparameters simultaneously: each polyline corresponds to a specific combination of hyperparameter values—namely, maximum number of blocks, inner tilewidth, and threads per block—and is color-coded to reflect its relative runtime compared to all other configurations (with yellow indicating faster performance and blue indicating slower). The plots clearly illustrate the critical role of hyperparameter tuning across both hardware platforms and numeric precisions. Among the three parameters, the inner tilewidth emerges as the dominant performance factor. This parameter determines how much data is kept in register space during execution, and its optimal value is inherently tied to the sizes of the L1 and L2 caches. For single precision, the optimal inner tilewidth is consistently 32, which aligns with a full 128-byte cache line; for double precision, the optimal value is 16, which also matches a full cache line at that precision. 

In addition to tilewidth, the importance of the other parameters shifts with the bandwidth regime. For bandwidth 32, performance is more sensitive to the maximum number of blocks, as this governs how many concurrent bulge-chasing operations can be scheduled and pipelined across execution units. On the other hand, for bandwidth 128, the number of threads per block becomes more critical. This is because each Householder reflector must touch a larger number of rows or columns, thus increasing intra-block parallelism requirements. Sufficient threads per block are necessary to expose and exploit this additional parallelism. Overall, across all GPU architectures (NVIDIA H100 and AMD MI300X) and all data precisions (FP32 and FP64), we consistently observe that larger values of both max blocks and threads per block tend to yield better performance—provided that the inner tilewidth is set to its architecture-aware optimal value. These findings underscore the importance of hardware-specific and precision-specific kernel tuning for achieving high efficiency in memory-bound GPU algorithms, in line with earlier findings that underline its importance\cite{ringoot2025performant,10.1007/978-3-031-69577-3_7}.

\subsection{Impact of Hardware Evolution on Performance}
\begin{figure}[t]
\centerline{\includegraphics[width=\linewidth]{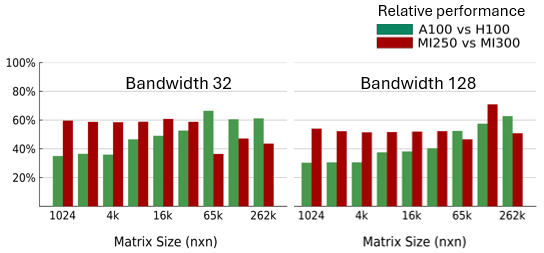}}
\caption{Performance gains from GPU architectural advancements, showing the relative performance loss of using older GPU architecture for varying matrix sizes and bandwidths across AMD and NVIDIA, and highlighting the impact of larger L1 and L2 caches.}
\label{fig:arch}
\end{figure}

Figure~\ref{fig:arch} illustrates the relative performance gains achieved by transitioning from older to newer GPU architectures—specifically, from MI250X to MI300X and from A100 to H100—across varying matrix sizes and bandwidths. The MI300X shows substantial improvement over the MI250X, which we attribute to its doubled L1 cache and the addition of a unified L2.5 (Infinity Cache) layer (see Table~\ref{tab:hardware}). Similarly, the H100 consistently outperforms the A100, benefiting from a 33\% increase in L1 cache and a 25\% boost in L2 capacity. These results reinforce our earlier profiling analysis (Section~\ref{sect:prof}) that L1 and L2 cache size and reuse are critical factors for high performance in memory-bound kernels.

\begin{figure}[t]
\centerline{\includegraphics[width=\linewidth]{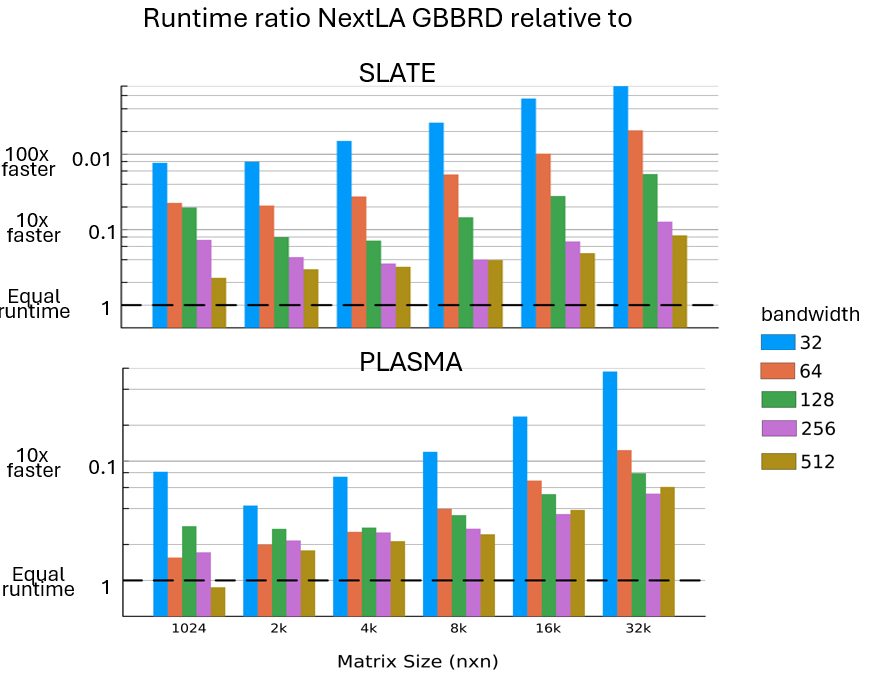}}
\caption{Runtime ratio of the GPU-accelerated band-to-bidiagonal reduction (GBBRD) relative to CPU-based implementations in SLATE (top) and PLASMA (bottom), across matrix sizes (1024–32k) and bandwidths (32–512). Values above the dashed horizontal line indicate configurations where the GPU implementation outperforms the CPU baseline. The strongest speed—ups to $60\times$  over PLASMA and $800\times$  over SLATE—are observed for small bandwidths and large matrices, highlighting the efficiency of the GPU-centric design even in previously CPU-favored regimes.}
\label{fig:slate}
\end{figure}

%\textcolor{red}{TODO: profile AMD+A200}

\subsection{State-of-the-Art Library Performance Comparison}
We evaluate the performance of our GPU-based band-to-bidiagonal reduction algorithm across a wide range of problem sizes and bandwidths, as depicted in Figure~\ref{fig:slate}. Specifically, we benchmark matrix sizes from $n = 1024$ to $n = 32{,}768$, and bandwidths ranging from 32 to 512, and compare against CPU-based implementations from two leading high-performance computing libraries: SLATE and PLASMA. \textcolor{black}{The optimized hyperparameters selected in Section \ref{sect:paramtune} are used in our implementation.} 
The results show that our GPU implementation consistently outperforms both libraries in nearly all tested configurations. At smaller bandwidths—such as 32, which are common in practical SVD scenarios—our algorithm achieves speed-ups between $4\times$ and $60\times$ over PLASMA and between $100\times$ and $800\times$ over SLATE. \textcolor{black}{While in certain cases the performance of the CPU libraries on multiple nodes could improve, we are comparing a single GPU with a single multicore CPU: the GPU algorithm could equally be extended to take advantage of multiple nodes.} These dramatic gains arise from our ability to fully exploit the GPU hierarchical massive parallelism and related memory hierarchy and apply fine-grained concurrency across multiple memory levels (registers, L1/L2 caches, and DRAM), as discussed in Sections~\ref{sect:prof} and~\ref{sect:occ}. 

Even at larger bandwidths, which are more demanding in terms of memory traffic, our implementation maintains significant speed-ups. For instance, at bandwidth 512, the GPU version achieves speed-ups of $2\times$ to $8\times$ over SLATE and $0.9\times$ to $6\times$ over PLASMA, depending on the matrix size. Interestingly, the GPU performance improves more rapidly with increasing matrix size than the CPU implementations. This trend aligns with the theoretical occupancy model outlined in Section~\ref{sect:occ}, where large problem sizes fully saturate GPU ALUs due to the tiling and spacing pattern of the bulge-chasing algorithm. These findings underscore the broader implication that the performance of GPU-resident algorithms—when optimized for memory locality, register pressure, and controlled occupancy—can rival and even significantly exceed mature CPU-based solutions in domains traditionally considered ill-suited for GPU acceleration. 

\begin{figure}[t]
\centerline{\includegraphics[width=\linewidth]{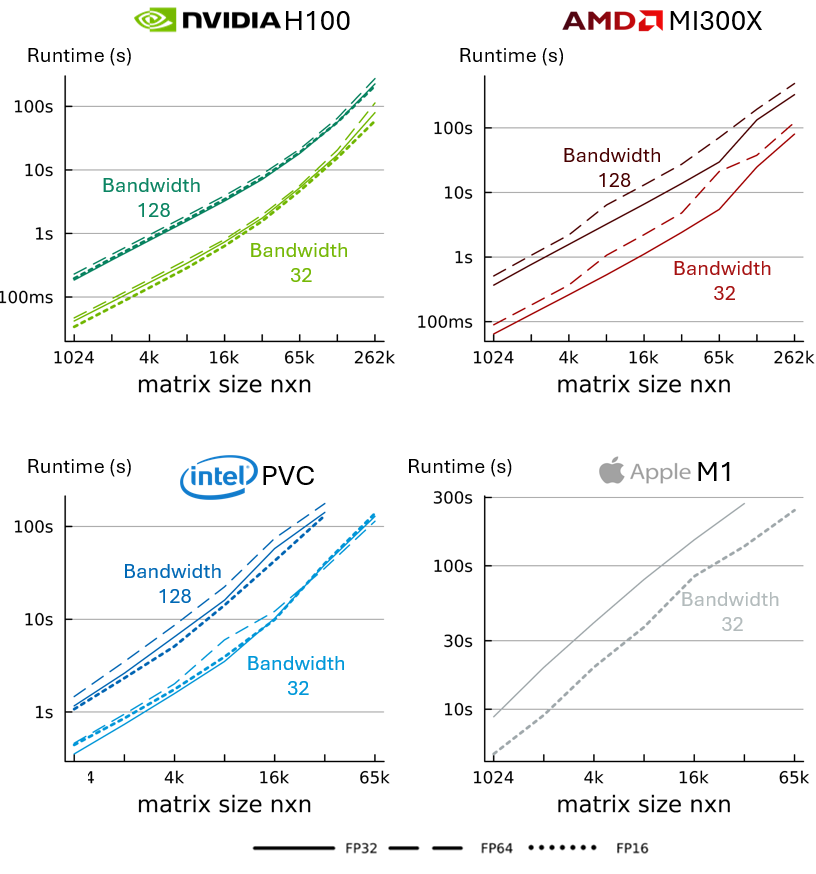}}
\caption{Runtime scaling of the reduction from banded to bidiagonal form across NVIDIA H100, AMD MI300X, Intel PVC, and Apple M1 for varying matrix sizes, bandwidths, and precisions, demonstrating portability and the power of abstractions.}
\label{fig:hardware}
\end{figure}
\subsection{Performance across hardware and data precision}
We implement the algorithm in a hardware- and precision-agnostic fashion: a single function definition is written and automatically specialized by the compiler for each target architecture and data type. Additionally, performance portability is achieved through user-level hyperparameter tuning, enabling algorithmic adaptation to device-specific characteristics. 
\textcolor{black}{A hardware-adapted suggestion of hyperparameter is provided to end-users, but they can directly tune the parameters as desired. }
Figure~\ref{fig:hardware} illustrates the cross-platform portability and performance of the implementation on GPUs from NVIDIA (H100), AMD (MI300X), Intel (PVC 1100), and Apple (M1), evaluated at bandwidths 32 and 128 for FP16, FP32, and FP64 data types. While all architectures successfully run the same high-level code, the achieved performance varies. AMD’s MI300X exhibits a modest slowdown of approximately 1.5–2$\times$ compared to NVIDIA’s H100. In contrast, Intel’s PVC shows a performance degradation of roughly 20$\times$, despite its larger cache sizes (see Table~\ref{tab:hardware}). This discrepancy highlights a key insight: cache size alone is not a strong predictor of performance. Rather, L1 and L2 bandwidth, which is strongly linked to their latency, is a more critical bottleneck. The observed performance ranking across architectures aligns closely with their reported L1/L2 access latencies, emphasizing the importance of fast memory access in addition to absolute cache capacity in memory-bound GPU workloads. \textcolor{black}{Indeed, while our algorithm leverages hyperparameter tuning to adapt to various architectures and data precisions, its performance ultimately remains constrained by hardware-specific characteristics.} In the user library, a heuristic per architecture can be provided, as well as the option for the user to fine-tune the parameters themselves for novel architectures with one additional line of code, leveraging the Julia Language multiple-dispatch feature.

\section{Conclusion}
This work redefines the performance boundaries of memory-bound linear algebra on modern GPUs. By designing the first GPU-resident memory-aware band-to-bidiagonal reduction algorithm—backed by a tile-wise strategy and implemented in portable Julia—we demonstrate that GPU architectures can not only match but decisively surpass CPU performance in domains historically considered off-limits to accelerators. Our findings show that optimal performance emerges not from hardware-specific rewrites, but from principled algorithm design, cache-aware parameterization, and portable compiler infrastructure. Our implementation leverages Julia’s high-level abstractions 
to deliver a single-source codebase that is portable across GPU vendors and data precisions. We validate this on NVIDIA, AMD, Intel, and Apple Metal hardware, covering FP16, FP32, and FP64. Performance scales robustly across precisions, with error analysis showing minimal loss in accuracy even at wide matrix bandwidths and large matrix sizes.

Through benchmarking, we demonstrate that our GPU implementation outperforms state-of-the-art libraries (PLASMA and SLATE) starting from matrix sizes as small as $1024\times1024$, achieving speed-ups up to 800× for $32k \times 32k$ matrices at narrow bandwidths. In addition, across all architectures, we find that performance is not solely dictated by DRAM throughput, but hinges critically on L1/L2 memory. Finally, our hardware study underscores that L1/L2 cache bandwidth, not just size, is the key limiting performance factor. Despite Intel PVC offering higher cache capacity than AMD or NVIDIA, its higher L1/L2 latency results in substantially lower performance, emphasizing the importance of low-latency fast memory access in memory-bound workloads, and indicating the importance of algorithmic considerations when developing future hardware generations. Overall, our results demonstrate that GPU architectures—when paired with careful hyperparameter tuning and architecture-aware design—can now support high-performance, accurate reduction from banded form to bidiagonal form for singular value computation.

\section{Future Work}
In this work, we presented a hardware-agnostic GPU kernel implementation for reducing banded matrices to bidiagonal form, achieving high performance across GPU architectures through hyperparameter tuning. Future work could integrate auto-tuning approaches~\cite{10.1007/978-3-031-69577-3_7} to eliminate manual tuning and enable adaptation to new hardware. To scale further across mixed heterogeneous environments, task-based parallelism and dynamic dataflow execution could be managed using Dagger.jl~\cite{alomairydynamic}, enabling performance-portable scheduling across devices. Most importantly, this work bridges a missing link in the full GPU-resident SVD pipeline. While singular value computation for banded matrices is relevant in direct applications~\cite{meuris2023machine}, it also forms a key stage in the broader SVD process. Hardware-agnostic kernels have been developed for the first stage (reduction to banded form)~\cite{ringoot2025performant}, and others have implemented the third stage (bidiagonal diagonalization) on GPUs~\cite{laszlo2016manycore,tolmachev2025high,miao2015fast}. By accelerating the second stage, our work shifts the trade-off between bandwidth and runtime, potentially enabling faster end-to-end SVD computations. Finally, extending this algorithm to compute singular vectors—similar to prior work on GPU-resident eigenvector computation~\cite{10.1145/3721145.3730411,10.1145/3710848.3710894}—could enable a fully GPU-resident, hardware-agnostic full SVD. This positions our work as a key step toward efficient, accelerator-native numerical linear algebra pipelines.

\section*{Acknowledgments}
We are grateful for the work and support of members of the Julia Lab, in particular Valentin Churavy, Tim Besard, James Schloss, and Julian Samaroo, who developed and helped with our understanding of available tools within the Julia ecosystem for performance portability. R.A. acknowledges the KAUST Ibn Rushd post-doctoral fellowship. The authors acknowledge the MIT Office of Research Computing and Data, ACES at Texas A and M HPRC (CIS250776 ACCESS,  supported by U.S. NSF 2138259, 2138286, 2138307, 2137603, 2138296), and the Advanced Micro Devices Developer Cloud for providing high-performance computing resources. This material is based upon work supported by the US NSF (CNS-2346520, PHY-2028125, RISE-2425761, DMS-2325184, OAC-2103804, OSI-2029670), DARPA (HR00112490488), DoE (DE-NA0003965), and USAFR (FA8750-19-2-1000). The U.S. Government, its agencies, and their employees do not make any warranty, do not endorse, recommend, or favor, nor assume any liability for anything in this report, nor represent that its use would not infringe privately owned rights. The views and opinions expressed herein are those of the authors only.

\bibliographystyle{IEEEtran}
\bibliography{bib}

\end{document}